\documentclass[12pt]{iopart}

\usepackage{braket}
\usepackage{graphicx}
\usepackage{units}
\usepackage[colorlinks=true,citecolor=blue,urlcolor=blue,linkcolor=blue]{hyperref}
\usepackage{bbold}
\usepackage{xcolor}
\usepackage{multirow}

\usepackage{booktabs}
\usepackage{array}

\graphicspath{{../figs/}}
\usepackage{subfig}
\usepackage{amssymb}
\usepackage{ulem}
\usepackage{amsfonts}
\usepackage{mathrsfs}
\usepackage{dcolumn} 
\usepackage{bm}
\usepackage{color}
\usepackage{cancel}
\usepackage{mathbbol}
\usepackage{epsfig}
\usepackage{units}
\usepackage{esint}
\usepackage{soul} 
\usepackage{url}

\definecolor{rot}{rgb}{0.75,0.05,0.25}
\definecolor{hellgrau}{gray}{0.5}
\definecolor{blau}{rgb}{0,0,0.7}

\newcommand{\av}[1]{\langle#1\rangle}

\begin{document}

\title[]{Experimental test of fluctuation relations for driven open quantum systems with an NV center}

\author{S. Hern\'{a}ndez-G\'{o}mez$^{1,2,3}$, N. Staudenmaier$^{1,2}$\footnote{Current address: Institute for Quantum Optics, Ulm University, D-89081, Germany}, M. Campisi$^{4,2,5}$, N. Fabbri$^{1,3}$}
\address{$^1$ European Laboratory for Non-linear Spectroscopy (LENS), Universit\`a di Firenze, I-50019 Sesto Fiorentino, Italy}
\address{$^2$Dipartimento di Fisica e Astronomia, Universit\`a di Firenze, I-50019, Sesto Fiorentino, Italy}
\address{$^3$Istituto Nazionale di Ottica del Consiglio Nazionale delle Ricerche (CNR-INO), I-50019 Sesto Fiorentino, Italy}
\address{$^4$NEST, Istituto Nanoscienze-CNR and Scuola Normale Superiore, I-56127 Pisa, Italy}
\address{$^5$INFN - Sezione di Pisa, I-56127 Pisa, Italy}
\ead{fabbri@lens.unifi.it}
\vspace{10pt}

\begin{abstract} 
The experimental verification of quantum fluctuation relations for driven open quantum system is currently a challenge, due to the conceptual and operative difficulty of distinguishing work and heat. The Nitrogen-Vacancy center in diamond has been recently proposed as a controlled test bed to study fluctuation relations in the presence of an engineered dissipative channel, in absence of work [Hern\'andez-G\'omez et al, Phys. Rev. Research 2, 023327 (2020)]. Here, we extend those studies to exploring the validity of quantum fluctuation relations in a driven-dissipative scenario, where the spin exchanges energy both with its surroundings because of a thermal gradient, and with an external work source. We experimentally prove the validity of the quantum fluctuation relations in the presence of cyclic driving in two cases, when the spin exchanges energy with an effective infinite-temperature reservoir, and when the total work vanishes at stroboscopic times -- although the power delivered to the NV center is non-null. Our results represent the first experimental study of quantum fluctuation relation in driven open quantum systems.
\end{abstract}

\section{Introduction}

Quantum fluctuation relations~\cite{Bochkov77SPJETP45,Bochkov79SPJETP49,Esposito09,Campisi11} are exact relations in  quantum statistical mechanics that allow for a compact and effective description of a plethora of non-equilibrium phenomena such as the statistical nature of the second law of thermodynamics and the emergence of an arrow  time~\cite{Jarzynski11ARCMP2}, Onsager relations, Green-Kubo relations and high order relations between responses and multi-point correlation functions~\cite{Andrieux08PRL100,Andrieux09}, the microscopic mechanisms at the basis of Clausius inequality~\cite{Jarzynski99JSM96}, and the Carnot theorem imposing the celebrated Carnot efficiency as the universal upper bound to efficiency of heat engines~\cite{Campisi14JPA47,Campisi15NJP17}. Recently, they have been shown to be at the basis of thermodynamic uncertainty relations expressing lower bounds on entropy production in terms of the precision (signal-to-noise ratio) of fluctuating non-equilibrium charges (i.e., integrated non-equilibrium currents)\cite{Barato15PRL114,Horowitz19NP16}, or the asymmetry of their statistics~\cite{Campisi2011.01076arXiv}.

Generally speaking, quantum fluctuation relations refer to a scenario where a quantum system, prepared in thermal equilibrium at some inverse temperature $\beta_1$, is then placed in contact (sequentially or at the same time) with several thermal  baths at various temperatures $\beta_2, \dots \beta_N$ while being possibly acted upon by a work source resulting in a  time-dependent forcing~\cite{Jarzynski99JSM96}. In integral form, the quantum fluctuation relation reads~\cite{Campisi11}:
\begin{equation}
\av{e^{-\sum_i \beta_i \Delta E_i} }= e^{-\sum_i \beta_i \Delta F_i}
\label{eq:generalFT}
\end{equation}
where $\Delta F_i$ is the free energy change between the (hypothetical) equilibrium state of component $i$ of the global system+bath compound at the protocol's end time, and its initial free energy, while $\Delta E_i$ is the according stochastic energy change (here the so called two-point measurement scheme is implied~\cite{Campisi11}). The symbol $\av{\cdot}$ denotes average over the statistics of the joint probability distribution $p(\Delta E_1, \dots \Delta E_N)$ for the energy change of each component.

The above relation encompasses a number of typical situations encountered in non-equilibrium thermodynamics. 
For example, a system initially in equilibrium and in contact with a thermal bath at some inverse temperature $\beta$ is acted upon by a time dependent force: In this case $\beta_1 = \beta_2=\beta$, and since the work done by the external source is given by $W=\Delta E_1+\Delta E_2$, the Jarzynski relation follows $\av{e^{-\beta W}}=e^{-\beta \Delta F}$, with $\Delta F$ the global free energy variation~\cite{Campisi09}. While this relation has been verified experimentally with closed quantum systems, (namely in absence of a thermal bath)~\cite{An15,Batalhao14}, its verification with an open quantum system is still elusive, due to the fact that the measurement of total energy injected in system and bath is typically very challenging. Possible solutions to this challenge are offered by the interferometric method for the measurement of work statistics~\cite{Dorner13PRL110,Mazzola13PRL110} as discussed in~\cite{Campisi13NJP15}, or with reference to superconducting circuit setups, by calorimetric measurements of energy exchanges between a qubit and a resistor~\cite{Hekking13PRL111}.

A complementary situation is when two quantum systems prepared initially at different temperatures, are placed in thermal contact for some time $\tau$. Provided the interaction energy is negligible, no work is associated to this process, and the general relation takes the form of the so called exchange fluctuation relation $\av{e^{\Delta \beta \Delta E}}=1$ 
\cite{Jarzynski04}, where $\Delta \beta = \beta_2 -\beta_1$ and $\Delta E = \Delta E_1 = -\Delta E_2$. This relation has been recently verified in the quantum regime with NMR setup ~\cite{Pal19PRA100} and with NV centers in diamond~\cite{HernandezGomez20}.

A situation of central interest in quantum thermodynamics combines the previous two scenarios, namely when not only a system exchanges energy with its surroundings because of a thermal gradient, but also with an external work source. That is indeed what happens during the operation of a heat engine. The experimental verification of Eq.~(\ref{eq:generalFT}) when there is a simultaneous exchange of heat and work poses significant challenges, as typical quantum platforms allow to experimentally access the system energy change $\Delta E=W-Q$, but not the individual terms $W$ and $Q$.

In this work, we address this challenge and experimentally explore the validity the quantum fluctuation relation for a driven-dissipative dynamics (Eq.~(\ref{eq:generalFT})) 
realized by the ground state of a single negatively-charged Nitrogen Vacancy (NV) center in diamond.
The NV center is well established as a biocompatible quantum sensor~\cite{Rondin14,Barry20,HernandezGomez21}, single-photon source~\cite{Sipahigil2012}, as well as building block for quantum registers~\cite{Childress13,Bradley19,Thiering20}, thanks to the coherent control of its ground state spin and optical addressability. The capability of coherently manipulating a two or three level spin system, and to engineer dissipative channels to mimic the thermal contact with a quantum reservoir, make the NV center also an excellent platform to explore quantum thermodynamics processes~\cite{Klatzow19,HernandezGomez20}.
Here, we use an NV center spin qubit to prove the validity of the fluctuation relation in Eq.~(\ref{eq:generalFT}) in two special cases of driven-dissipative dynamics, where the measurement of the spin energy change $\Delta E$ suffices: i) when the environment with which the system exchanges energy has infinite temperature; ii) when the duration of the process is chosen in such a way that the total work done on the system amounts to zero, although generally the power delivered to it is non-null.

The results obtained in this work expand the experimental basis over which quantum fluctuation relations have been tested, and prepare the ground for a fully fledged testing of Eq.~(\ref{eq:generalFT}) in the general case when energy is exchanged both in the form of heat and work.

\section{Experimental setup and protocol}
\label{sec:setup_and_protocol}

To engineer a driven-dissipative dynamics, we use the electronic spin ground state of a single NV center in diamond at room temperature.
An external magnetic field, $B=38.9$~mT, is aligned along the NV spin quantization axis to remove the degeneracy between the  spin projections $m_S = \pm 1$. One of these states, $\ket{m_S=+1}\equiv\ket{1}$, is selected to form the two-level system basis along with $\ket{m_S=0}\equiv\ket{0}$.
The two-level system is optically initialized into $\ket{0}$ before each experiment, and its state is read out by measuring the NV spin-selective photoluminescence intensity~\cite{Dobrovitski13,Doherty13}.

We apply a continuous resonant microwave (mw)  
driving field, described by a Hamiltonian that, in the mw rotating frame, reads
\begin{equation}
\mathcal{H}(t) = \frac{\hbar \omega(t)}{2} \left(\sigma_x\cos\theta t  + \sigma_y\sin\theta t  \right) ,
\label{eq:H_general}
\end{equation}
where $\sigma_i$ are Pauli matrices, $\omega(t)$ is the Rabi frequency that is set by the mw amplitude, and $\theta$ is an angular frequency of the mw phase variation. 
We denote the instantaneous eigenstates of the Hamiltonian in Eq.~(\ref{eq:H_general}) as $\rho_{\pm}{(t)}\equiv\ket{E_{\pm}{(t)}}\!\!\bra{E_{\pm}{(t)}}$, and their corresponding eigenvalues as $E_{\pm}{(t)} = \pm {\hbar \omega(t)}/{2} $. 

\begin{figure}
\begin{center}
\includegraphics[width=0.6\textwidth]{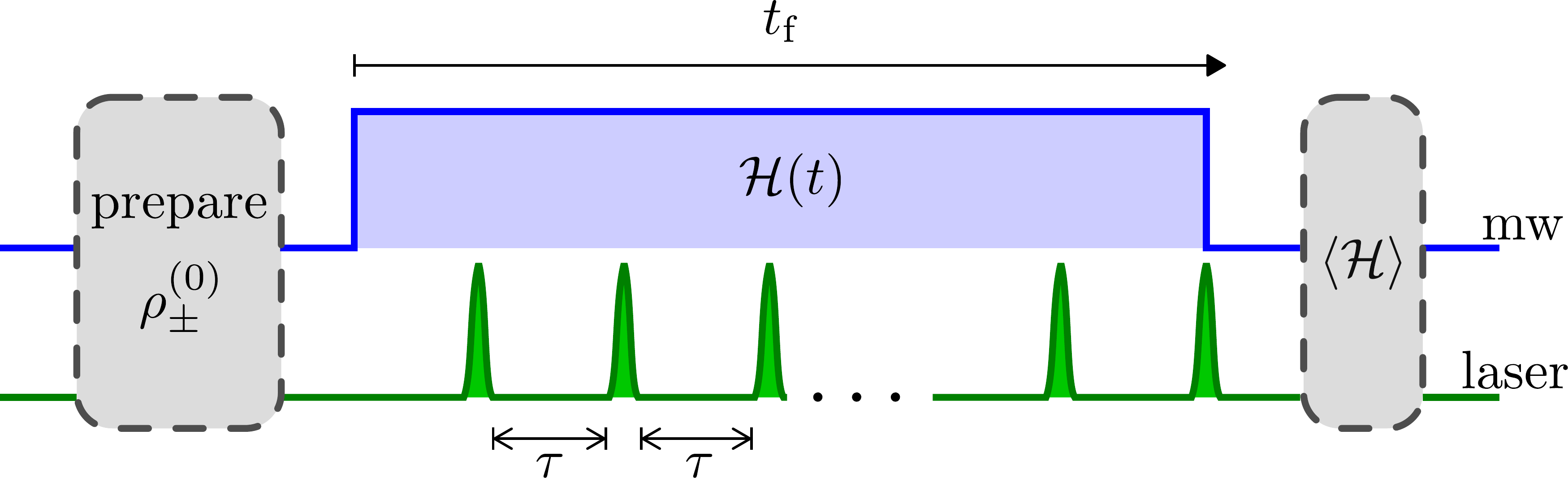}
\end{center}
\caption{Scheme of the experimental protocol. The combination of coherent  mw driving (blue) and a train of equidistant short laser pulses (green) gives rise to a dissipative map $\mathcal{M}$~(see text). The initially prepared eigenstates of the Hamiltonian are subject to this map, and the energy of the system is measured at the very end of the protocol. The time between pulses $\tau$ is fixed, 
so increasing the final time of the experiment $t_\mathrm{f}$ results in an increment on the number of short laser pulses $N_\mathrm{L}$ applied during the protocol.}
\label{fig:scheme_experiment}
\end{figure}

Besides the unitary evolution, the system interacts with a train of $N_\mathrm{L}$ temporally-equidistant short laser pulses, as depicted in Fig.~\ref{fig:scheme_experiment}. 
In a previous work~\cite{HernandezGomez20}, we have demonstrated that a photon absorption from each laser pulse results in the application of a quantum projective measurement and the opening of a dissipative channel. The application of a train of short laser pulses brings the spin into a asymptotic Gibbs state with inverse (pseudo)temperature $\beta_R$~\cite{HernandezGomez20} and, as such,
is  analogous to the dissipation channel caused by the interaction with a  heat reservoir $R$.
The value of $\beta_R$ is determined by a combination of factors, which are the relative orientation of the driving Hamiltonian operator with respect to the NV quantization axis ($\sigma_z$), the Rabi frequency of the driving field, the laser absorption probability $p_\mathrm{a}$, and the time between pulses~\cite{HernandezGomez20}. 

To measure the statistics of the spin energy change, and to verify the fluctuation relation in Eq.~(\ref{eq:generalFT}), we implement an experimental protocol  consisting of three steps~\cite{HernandezGomez20}: 
\begin{description}
\item(i)~{\it Initialization.} The system is initialized into one of the initial Hamiltonian eigenstates $\{\rho_{\pm}{(0)}\}$, say  $\{\rho_{+}{(0)}\}$, by applying a long laser pulse that optically pumps the spin in $\ket{0}$, followed by an opportune spin-rotating mw gate.
\item(ii)~{\it Evolution.} The system evolves during a time $t_{\mathrm{f}}$ under the action of the mw driving, and is intermittently perturbed by short laser pulses.
\item (iii)~{\it Readout.} At time $t_{\mathrm{f}}$, the spin is mapped back to the $\{\ket{0},\ket{1}\}$ basis and the NV photoluminescence intensity under laser illumination is recorded to measure the spin state. 
\end{description}

The protocol is repeated several times and the integrated photoluminescence signal gives the probability for the spin to end up in state $\rho_{+}{(t)}$. 
As a result, we experimentally construct the conditional probability $P_{+|+}{(t_{\mathrm{f}})}$ that the state $\rho_{+}{(t)}$ is detected at time $t=t_{\mathrm{f}}$ given that the two-level-system was initialized into $\rho_{+}{(0)}$ at time $t=0$.  The same procedure is applied with the system initialized into $\rho_-(0)$ to give $P_{+|-}{(t_{\mathrm{f}})}$.

Formally, the conditional probability is expressed as 
\begin{equation}
P_{+|\pm}{(t_{\mathrm{f}})}\equiv \mathrm{Tr} \left[ \rho_{+}{(t_{\mathrm{f}})} \,\mathcal{M}(t_{\mathrm{f}})[\rho_{\pm}{(0)}] \right]
\end{equation}
where $\mathcal{M}(t_{\mathrm{f}})[\rho_{\pm}{(0)}]$ is the density operator at time $t_{\mathrm{f}}$, expressed as a quantum map $\mathcal{M}(t_{\mathrm{f}})$ applied to the initial state $\rho_{\pm}{(0)}$. 
Notice that, by definition, $P_{-|\pm} = 1- P_{+|\pm}$. 
Therefore, measuring $P_{+|+}{(t_{\mathrm{f}})}$ and $P_{+|-}{(t_{\mathrm{f}})}$ is enough to completely reconstruct the energy jump probabilities. 

Together with the initial probability $P_i{(0)}$,  the conditional probability $P_{j|i}{(t_{\mathrm{f}})}$, with $j=\pm$ and $i=\pm$,  provides all necessary information to construct the two-level-system energy change probability distribution function: 
\begin{equation}\label{eq:pDeltaE}
p(\Delta E)= \sum_{i,j} \delta (\Delta E -E_j{(t_{\mathrm{f}})}+E_i{(0)}) P_{j|i}{(t_{\mathrm{f}})} P_i{(0)} .
\end{equation}
During the statistical analysis of the experimental data, the results for each of the initial states are weighted with the according Gibbs probability 
$P_{\pm}{(0)}=e^{-\beta E_{\pm}{(0)}}/Z_0$ hence emulating a projective measurement being performed onto a thermal mixture at inverse temperature $\beta$. 
Given that we measure the conditional probabilities $P_{j|i}{(t_{\mathrm{f}})}$ for a high number of experimental realizations ($\sim 10^6$) and for each Hamiltonian eigenstate, this protocol is equivalent to the two-time measurement protocol applied to a thermal mixed state with inverse temperature $\beta$. 

The NV center is an open quantum system: during its evolution it exchanges energy in the form of heat with the laser field (mimicking a thermal reservoir) and in the form of work with the external time dependent driving. Using the symbol $Q$ to denote the energy gained by the reservoir, Eq.~(\ref{eq:generalFT}) becomes in the present case:
\begin{equation}
\av{e^{-\beta \Delta E - \beta_R Q}}= e^{-\beta \Delta F}
\label{eq:FT}
\end{equation}
where $\beta_R$ denotes the reservoir inverse (pseudo)temperature, and $\Delta F= -\beta^{-1}\ln Z(t_{\mathrm{f}})/Z(0)$ is the free energy difference of the NV center, between its reference equilibrium state at time $t_{\mathrm{f}}$ and its initial equilibrium state. Note that the external driving only applies to the NV center and does not affect the reservoir, therefore the reservoir free energy is constant ($\Delta F_R=0$) and hence it does not appear in the above equation. In Eq.~(\ref{eq:FT}) the symbol $\av{\cdot}$ denotes the average over the joint probability distribution function $p(\Delta E, Q)$, while we only have experimental access to its marginal $p(\Delta E)$. So, in principle, in order to check the validity of Eq.~(\ref{eq:FT}) one needs to perform measurements of both $Q$ and $\Delta E$. The marginal distribution $p(\Delta E)$ is however sufficient to check the validity of Eq.~(\ref{eq:generalFT}) in two special cases, namely when the reservoir temperature is infinite ($\beta_R=0$), and when the total work $W= \Delta E+ Q$ performed on the system is null, and accordingly $\Delta E= -Q$.

In the first case Eq.~(\ref{eq:FT}) reduces to:
\begin{equation}\label{eq:FTa}
\av{e^{-\beta \Delta E }}= e^{-\beta \Delta F}.
\end{equation}
Since the heat $Q$ does not explicitely appear in Eq.~(\ref{eq:FTa}), the average reduces to the one over the marginal $p(\Delta E)$. Note that although this expression appears formally similar to the Jarzynski relation~\cite{Jarzynski97}, in an open system as considered here the energy exchange takes into account not only work but also heat flux. 
In the second case, Eq.~(\ref{eq:FT}) reduces to:
\begin{equation}
\av{e^{-(\beta-\beta_R) \Delta E }}= e^{-\beta \Delta F}
\label{eq:FTb}
\end{equation}
where, for the same reason as above, the average can be understood as an average over the marginal $p(\Delta E)$. 
Note that in our experiment the power delivered to the NV spin is non-null at all times, namely the NV spin continuously exchanges energy with the work source. However, we will focus on specific times $t_{\mathrm{f}}$ at which the total work amounts to zero, in which case Eq.~(\ref{eq:FTb}) holds.

\section{Reservoir at infinite pseudo-temperature}
\label{sec:cyclicAmpVar}

\begin{figure}
\centering
\subfloat{\includegraphics[width=0.85\textwidth]{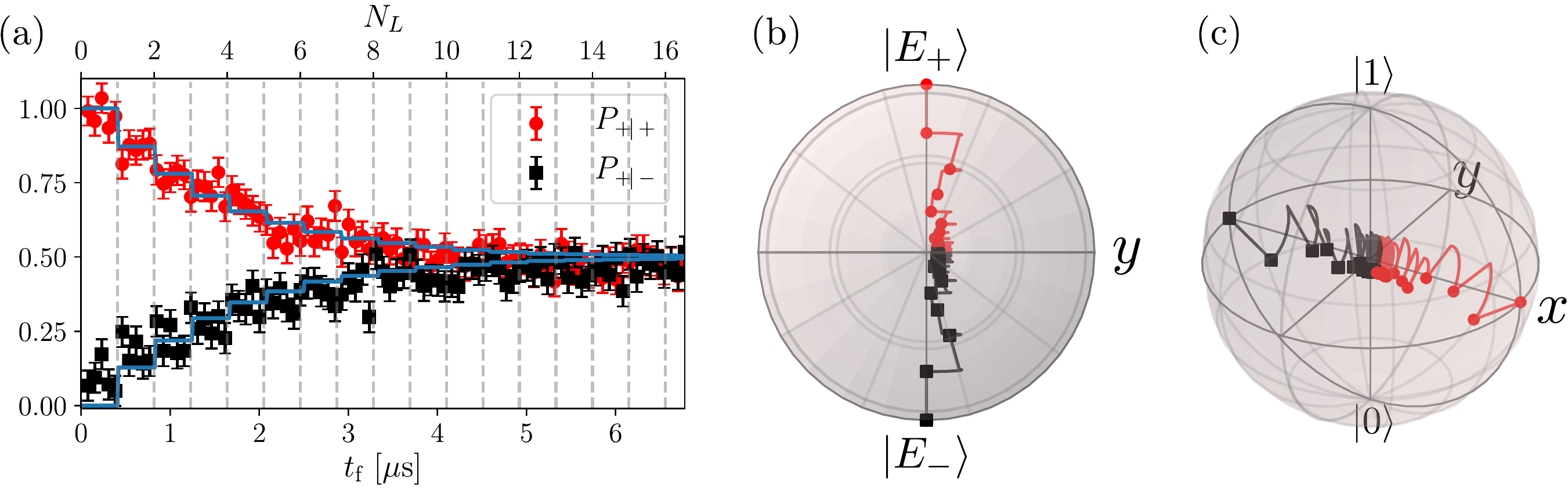}}
\caption{\textbf{(a)} Conditional probability of the spin to be measured in the eigenstate $\ket{E_+}$, after initialization either in $\ket{E_+}$ (red dots) or in $\ket{E_-}$ (black squares), reported as a function of time $t_{\mathrm{f}}$, for a Hamiltonian aligned along $\sigma_x$  with time-varying amplitude.
The time between laser pulses is $\tau=410$~ns. The top axis reports the time when the $N_\mathrm{L}$-th laser pulse is applied.
Blue line: numerical simulation, as detailed in~\ref{app:num_model}.
Notice how the conditional probabilities only change at times when a laser pulse is applied. For long evolution time the probabilities asymptotically approach to a constant level equal to one half. 
\textbf{(b-c)} Bloch sphere showing simulations of the mean trajectories starting from each of the eigenstates $\ket{E_\pm}$, respectively shown in red and black (top \textbf{(b)} and front \textbf{(c)} view).
Each marker represents the mean state at times equal to multiples of $\tau$, immediately after interacting with a laser pulse.
The solid line between two consecutive markers shows two different kinds of dynamics, one associated with the coherent evolution around the $x$-axis, and a second one associated with the dissipation that pushes the states towards the center of the Bloch sphere, due to the presence of laser pulses.
The blue lines in \textbf{(a)} are the result of projecting the trajectories in \textbf{(b)} (or \textbf{(c)}) onto the Hamiltonian axis.}
\label{fig:condit_probs_amplitude}
\end{figure}

In this section we report on the experimental verification of Eq.~(\ref{eq:FTa}) as a special case of the general fluctuation relation in Eq.~(\ref{eq:generalFT}).
To that end we set an infinite pseudo-temperature of the reservoir that mimics the thermal environment ($\beta_R=0$).
That is achieved by applying the following Hamiltonian: 
\begin{equation} \label{eq:H_variable-amplitude}
\mathcal{H}(t) = \frac{\hbar \omega(t)}{2} \sigma_x.
\end{equation}
Applying a sequence of short laser pulses while the system evolves under this Hamiltonian results in our NV center qubit reaching the completely mixed state.
Work is provided via a periodic temporal variation of the Hamiltonian amplitude $\omega(t)$, which in takes the form:
\begin{equation} \label{eq:omega_t}
\omega(t) = \frac{\omega_0}{2} \left(1 + \cos^2\frac{\pi t}{\tau_A} \right), 
\end{equation}
with period set to $\tau_A=616$~ns, and $\omega_0=\pi/\tau_A$.
Under these conditions, the eigenstates of the Hamiltonian remain constant in time, while the eigenvalues are time-dependent. 

\begin{figure}
\centering
\includegraphics[width=1\textwidth]{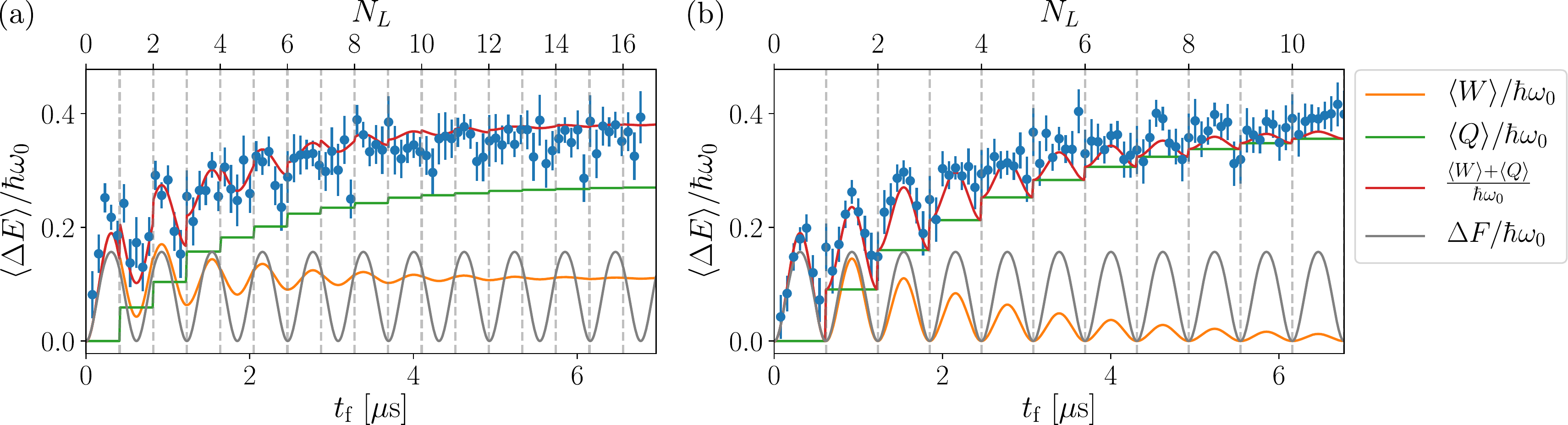}
\caption{
Experimental value of the mean energy variation $\av{\Delta E}$ (blue dots) for a Hamiltonian with time-varying amplitude. The time between consecutive laser pulses is set to \textbf{(a)}~$\tau=410\,\mathrm{ns}$, and \textbf{(b)}~$\tau=616\,\mathrm{ns}$, respectively.
Orange line: mean work $\av{W}$ applied to the qubit; green line: mean heat flux $\av{Q}$ exchanged with the reservoir at infinite pseudo-temperature.
Their sum $\av{W}+\av{Q}$ (red line) is in agreement with the experimental values of the mean energy variation. 
The free energy variation (gray line) only depends on the amplitude of the Hamiltonian (see text). 
The vertical dashed lines indicate the times at which laser pulses were applied to the spin qubit.}
\label{fig:DeltaE_cyclic_amplitude}
\end{figure}

\begin{figure}
\centering
\includegraphics[width=0.85\textwidth]{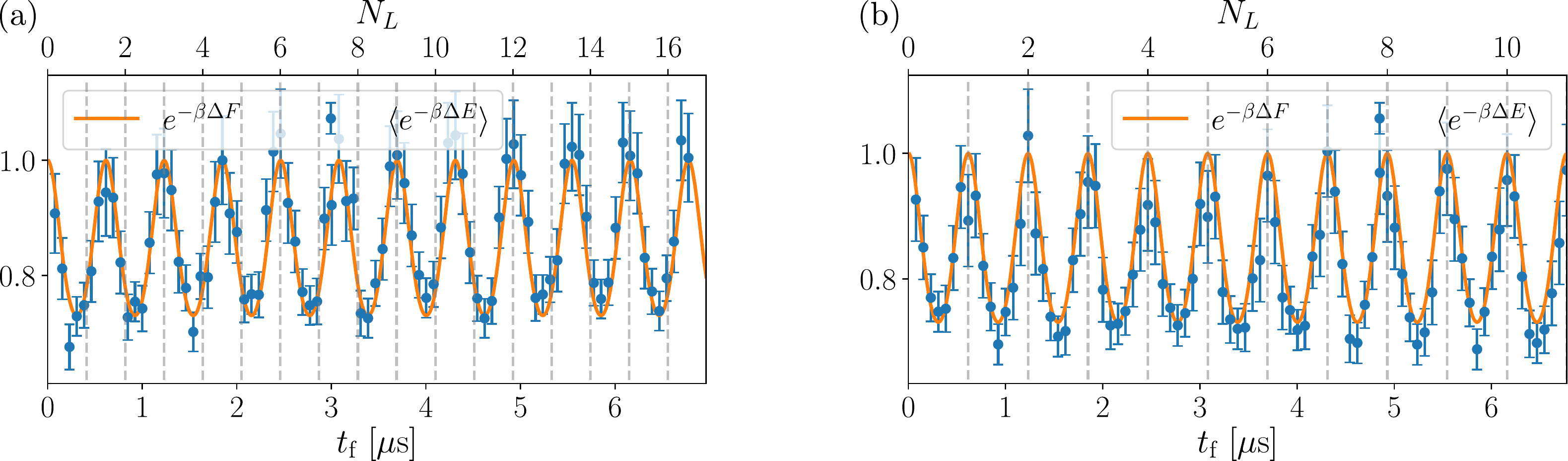}
\caption{
Experimental verification of the quantum fluctuation relation for a driven qubit in contact with a reservoir at infinite pseudo-temperature [Eq.~(\ref{eq:FTa})].
 \textbf{(a)}~$\tau=410\,\mathrm{ns}$, and \textbf{(b)}~$\tau=616\,\mathrm{ns}$.
Blue dots:  experimental values of the average exponentiated energy variation. Orange line: calculated $e^{-\beta \Delta F}$ (see text).}
\label{fig:FRHE_cyclic_amplitude}
\end{figure}

We applied the protocol described in the previous section to measure the conditional probabilities $P_{j|i}{(t_{\mathrm{f}})}$, considered as functions of $t_{\mathrm{f}}$.
Since the Hamiltonian commutes with itself at different time values, $P_{j|i}{(t_{\mathrm{f}})}$ remain constant during the time in-between laser pulses, while they present discontinuities caused by the dissipation at the times when laser pulses are applied, as shown in Fig.~\ref{fig:condit_probs_amplitude}(a). 
Each experimental point shown in this figure was obtained by averaging over an ensemble of~$\sim 10^6$ repetitions of the same protocol.

The asymptotic behavior is such that, in the limit of $t_{\mathrm{f}}\rightarrow\infty$, the conditional probabilities $P_{j|i}{(t_{\mathrm{f}}\rightarrow\infty)} \!\rightarrow 0.5$ for any $i$ and $j$, 
meaning that our dissipative channel is for practical purposes analogous to a thermal reservoir of inverse temperature  $\beta_R=0$, as anticipated above.
In Fig.~\ref{fig:condit_probs_amplitude}(b)-(c), we show the simulated mean trajectory obtained by propagating the density operators $\ket{E_{+}}\!\!\bra{E_{+}}$, and $\ket{E_{-}}\!\!\bra{E_{-}}$, with the dissipative map $\mathcal{M}$ that describes the dynamics.  See~\ref{app:num_model} for details on the simulation. 
The concatenation of coherent driving and dissipation results in a map that intermittently pushes the state towards the center of the Bloch sphere.

Having measured the conditional probabilities of energy variation, we can obtain the energy variation probability distribution function in Eq.~(\ref{eq:pDeltaE}). This is enough to reconstruct the mean value of a generic function of $\Delta E$. 

The mean value of the energy variation $\av{\Delta E}$ is shown in Fig.~\ref{fig:DeltaE_cyclic_amplitude}.
The probability associated with the initial Gibbs state was set to $P_+(0)=1/(1+e^2)$, hence the initial inverse temperature was $\beta = 2/\hbar\omega_0$.  
The experimental values retrieved for $\av{\Delta E}$ are in agreement with the theoretical values of the sum of the mean work applied on the system plus the mean heat flux, $\av{W}+\av{Q}$, as expected from the first law of thermodynamics. 
These mean values are calculated as $\av{W} = \int_0^{t_\mathrm{f}} \mathrm{d}t' \mathrm{Tr} [\rho(t')\frac{\mathrm{d}\mathcal{H}(t')}{\mathrm{d}t'}]$, and $\av{Q} = \int_0^{t_\mathrm{f}} \mathrm{d}t' \mathrm{Tr} [\frac{\mathrm{d}\rho(t')}{\mathrm{d}t'} \mathcal{H}(t')]$~\cite{Alicki79}. 
More details on this calculation can be found in~\ref{app:mean_W_and_Q}. 

In the case considered here, the Hamiltonian time dependency is reflected in the oscillations of $\av{W}$. 
On the other hand, the dissipation induced by the interaction with the laser pulses appears as discrete jumps in $\av{Q}$. 
As mentioned before, the dissipation tends to equilibrate the population of the Hamiltonian eigenstates, thus the energy variation induced by the Hamiltonian driving is gradually reduced, which explains the damping on $\av{W}$. 
For a time between laser pulses $\tau = \tau_A$ carefully chosen to be equal to the Hamiltonian period~[Fig.~\ref{fig:DeltaE_cyclic_amplitude}(b)], the flow of heat is perfectly synchronized to the times when work vanishes.
Instead, when the synchronization is not perfect $\tau \neq \tau_A$~[Fig.~\ref{fig:DeltaE_cyclic_amplitude}(a)], the mean work accumulates and tends to a finite value. 
In these experiments, the Hamiltonian is such that $Z(t) \equiv \mathrm{Tr} \left[e^{-\beta\mathcal{H}(t)}\right] = 2\cosh [\beta\ \hbar\omega(t)/2]$, with $\omega(t)$ defined in Eq.~(\ref{eq:omega_t}). 
Therefore, the values of $\Delta F = - \beta^{-1} \ln(Z(t_{\mathrm{f}})/{Z(0)})$ can be computed analytically (gray lines in Fig.~\ref{fig:DeltaE_cyclic_amplitude}). 
The periodic time-variation of the Hamiltonian induces an oscillation with constant amplitude on $\Delta F$. 
Notice that, before the very first laser pulse, when the system is closed, the difference $\av{W}_\mathrm{irr} \equiv \av{W}-\Delta F \geq 0$ is caused by the inner friction during the unitary evolution~\cite{Plastina14}. Given that the Hamiltonian eigenstates remain constant in time, the initial state remains unaltered by the unitary evolution. However, since the Hamiltonian eigenvalues change for a time $0<t_\mathrm{f}<\tau_A$, then the system is, on average, in a non-equilibrium state $\rho_{t_\mathrm{f}}=e^{-\beta\mathcal{H}(0)}/Z(0)$ (only describing an equilibrium thermal state for $t_\mathrm{f} = 0,\tau_A,2\tau_A,\dots$). 
The quantum relative entropy between $\rho_{t_\mathrm{f}}$ and the actual thermal state $\rho_{t_\mathrm{f}}^{\mathrm{th}} = e^{-\beta\mathcal{H}(t_\mathrm{f})}/Z(t_\mathrm{f})$ defines the value of $\av{W}_\mathrm{irr}$~\cite{Plastina14,Deffner10}. 
After the first laser pulse the system is no longer closed, and $\Delta F$ can exceed $\av{W}$. 
In contrast, the free energy variation is always smaller or equal than the mean energy variation, as expected from Jensen's inequality applied to Eq.~(\ref{eq:FTa}).

In order to effectively verify the validity of the FR in Eq.~(\ref{eq:FTa}), we used the experimental values of the energy variation probability distribution to obtain $\langle e^{-\beta \Delta E}\rangle$, and we compared it with the values of $e^{-\beta\Delta F} ={Z(t_{\mathrm{f}})}/{Z(0)}$ computed analytically, as shown in Fig.~\ref{fig:FRHE_cyclic_amplitude}. 
Notice that the oscillating $e^{-\beta\Delta F}$ depends only on the time-varying Hamiltonian, therefore, its values are the same in Fig.~\ref{fig:FRHE_cyclic_amplitude}(a), and Fig.~\ref{fig:FRHE_cyclic_amplitude}(b). 
In contrast, $\langle e^{-\beta \Delta E}\rangle$ depends on the time-varying Hamiltonian, and on the  energy variation probability distribution function, which in turns depends on the time between laser pulses, and the strength of the dissipation, i.e., the mean amplitude of the energy jumps induced by the interaction with a single laser pulse. 
The fact that work and heat flux are perfectly compensated to allow Eq.~(\ref{eq:FTa}) to be valid, is a consequence of the infinite pseudo-temperature of the environment. 
The agreement between the predicted values of $e^{-\beta\Delta F}$ and the experimental values of $\langle e^{-\beta \Delta E}\rangle$ verifies the validity of the FR in Eq.~(\ref{eq:FTa}). 

\section{Work vanishing at stroboscopic times}
\label{sec:cyclicPhase}
Now we consider a driving Hamiltonian with fixed amplitude $\omega(t) \equiv \omega_0$ and a time varying phase such that, 
\begin{equation}
\mathcal{H}(t) = \frac{\hbar \omega_0}{2} \left(\sigma_x\cos\theta t  + \sigma_y\sin\theta t \right) .
\label{eq:H_fixed-amplitude}
\end{equation}
In contrast to the case treated in the previous section, here the instantaneous eigenstates $\ket{E_\pm{(t)}}$ of $\mathcal{H}(t)$ are changing in time while the eigenvalues remain constant in time.  The Hamiltonian, and accordingly its eigenvectors, are periodic functions of time, with period $\tau_\theta= 2\pi /\theta$. As long as only the discrete times $t_n = n \tau_\theta$ are concerned, the discrete-time dynamics of the system, due to its periodicity, can be conveniently expressed in terms of an effective Floquet-type Hamiltonian, reading, in this case
\begin{equation} \label{eq:floquet_hamiltonian}
\mathcal{H}_{ \theta } = \frac{\hbar}{2} \left( \omega_0 \sigma_x + \theta \sigma_z \right) .
\end{equation}
The according unitary $U_\theta= e^{-i \mathcal{H}_{ \theta } \tau_\theta / \hbar}$ advances the system from time $n\tau_\theta$ to $(n+1)\tau_\theta$. As opposed to the original Hamiltonian $\mathcal{H}(t)$, the discrete-time Hamiltonian $\mathcal{H}_{ \theta }$ is time-independent.
Accordingly, in the discrete-time picture no work source appears to be at play.

\begin{figure}
\centering
\includegraphics[width=0.85\textwidth]{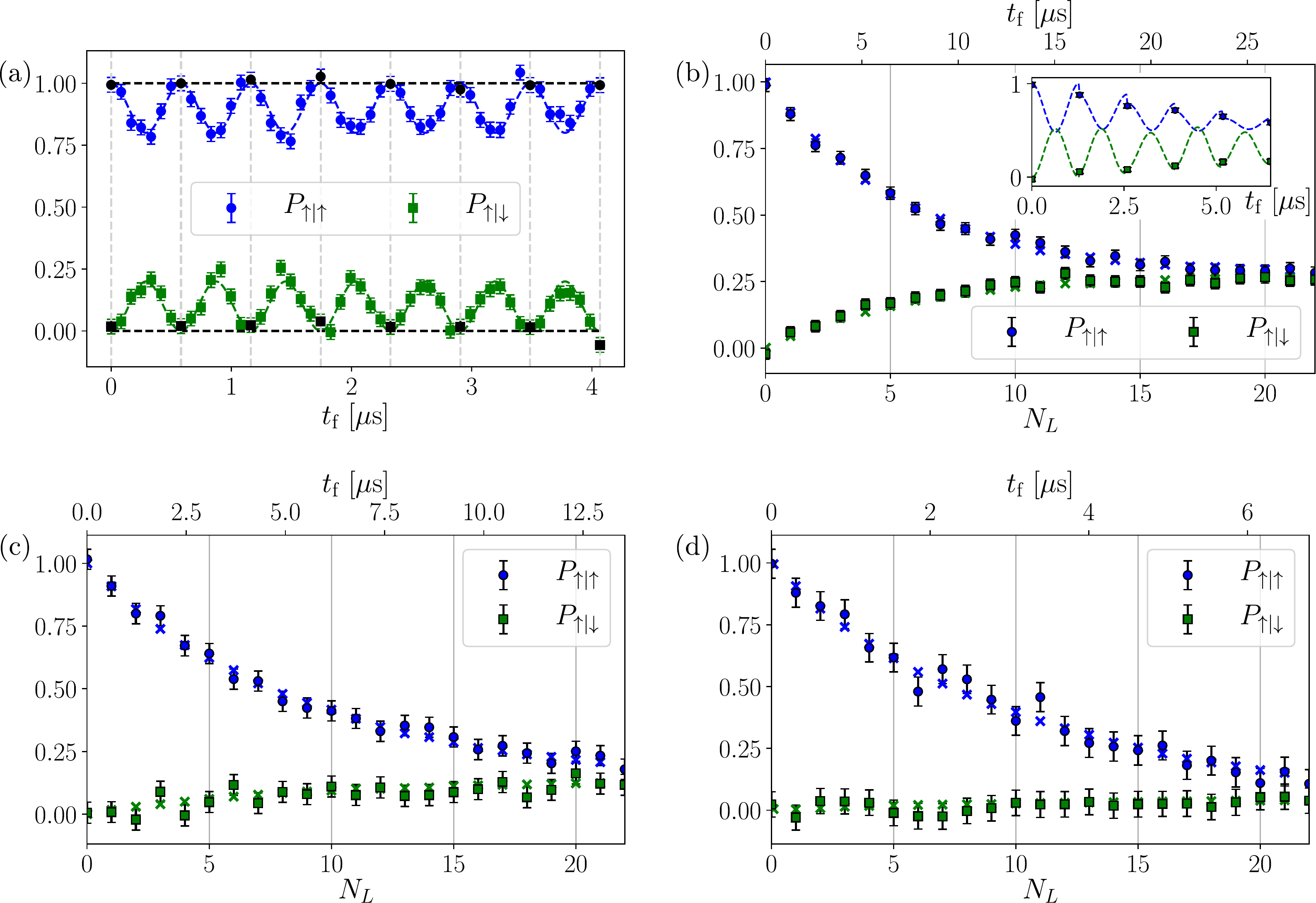}
\caption{Conditional probability for the spin qubit to be in the \textit{Floquet} eigenstate $\ket{E_{\theta,\uparrow}}$~[Eq.~(\ref{eq:floquet_eigenstate_up})], for an initial state $\ket{E_{\theta,\uparrow}}$ (blue circles) or $\ket{E_{\theta,\downarrow}}$ (green squares), in terms of time $t_\mathrm{f}$, for a Hamiltonian that rotates in the $xy$-plane with a period $\tau_\theta = 2\pi/\theta$.
{\bf (a)}~Without intermediate short laser pulses, with $\tau_\theta = 616\,\mathrm{ns}$, and $\alpha=\arctan(1/2)$. For synchronized stroboscopic measurements, the spin does not evolve in time (see black markers), as described in the main text. At intermediate times, the qubit performs Rabi oscillations.
{\bf (b-d)}~In the presence of short laser pulses, with interpulse spacing synchronized with $\tau_\theta$, as a function of time $t_\mathrm{f} = N_\mathrm{L} \tau_\theta$. 
\textbf{(b)} $\tau_\theta=1296\,\mathrm{ns}$, $\alpha=\pi/4$; \textbf{(c)} $\tau_\theta=616\,\mathrm{ns}$, $\alpha=\arctan(1/2)$; and \textbf{(d)} $\tau_\theta=308\,\mathrm{ns}$, $\alpha=\arctan(1/4)$.
For long times, the conditional probabilities approach a constant level $P_\uparrow^{\infty}$ that determines the reservoir inverse temperature. 
\textbf{(b)} $P_\uparrow^{\infty}=  0.276 \pm 0.003$; \textbf{(c)} $P_\uparrow^{\infty} =  0.138 \pm 0.002$; and \textbf{(d)} $P_\uparrow^{\infty} =  0.050 \pm 0.003$.
The crosses represent the result of Monte-Carlo simulations. 
Inset of \textbf{(b)}: Numerical simulation of the evolution between laser pulses.	
The bare Rabi frequency of the driving is in all these cases $\omega_0 \simeq (2\pi) 800$~kHz. 
}
\label{fig:condit_probs_phase}
\end{figure}

In our experiment the two-level system is prepared in one of the two eigenstates of $\mathcal{H}_{ \theta }$, reading
\begin{equation}
\ket{E_{\theta,\uparrow}} \equiv \sin(\alpha/2)\ket{1} + \cos(\alpha/2)\ket{0} \label{eq:floquet_eigenstate_up}
\end{equation}
\begin{equation}
\ket{E_{\theta,\downarrow}} \equiv \cos(\alpha/2)\ket{1} - \sin(\alpha/2)\ket{0} \label{eq:floquet_eigenstate_down}
 \end{equation}
where $\alpha = -\arctan\left({\omega_0}/{\theta}\right)$. 
The probability $P_{\uparrow|\uparrow}$ to start in the $\ket{\uparrow}$ state and find the system again in the $\ket{\uparrow}$ state after a time $t_{\mathrm{f}}$ reads $P_{\uparrow|\uparrow} = 1-\frac{\omega_0^2}{\omega_0^2 + \theta^2}\sin^2 \frac{t\theta}{2}$.
Similarly, it is $P_{\uparrow|\downarrow} = \frac{\omega_0^2}{\omega_0^2 + \theta^2}\sin^2 \frac{t\theta}{2}$. Note that these quantities are, as expected, periodic with period $\tau_\theta$. In addition, $P_{\uparrow|\uparrow}=1$ and  $P_{\downarrow|\uparrow} = 0$ at  stroboscopic times $t_n=n \tau_\theta$, in accordance with the fact that $\ket{E_{\theta,\uparrow}}$ and $\ket{E_{\theta,\downarrow}}$ are eigenvectors of the discrete-time Hamiltonian $\mathcal{H}_{ \theta }$. 
Fig. \ref{fig:condit_probs_phase}(a) shows the conditional probabilities $P_{\uparrow|\uparrow}$ and $P_{\uparrow|\downarrow}$  measured in absence of laser pulses. Note the agreement with the analytic expressions, confirming that $\mathcal{H}(t)$ well describes the dynamics in absence of laser pulses.

\begin{figure}
\centering
\includegraphics[width=\textwidth]{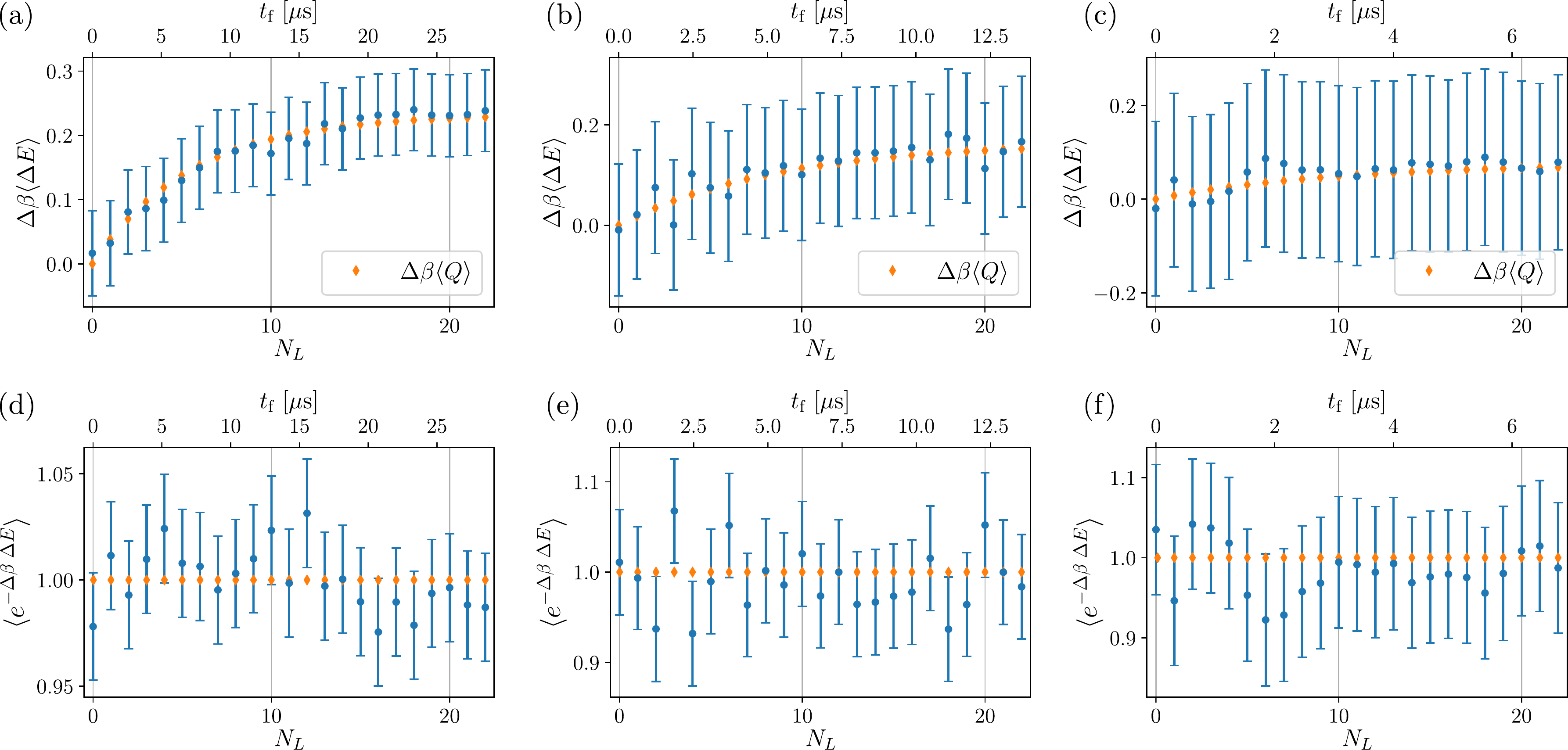}
\caption{
\textbf{(a-c)} Comparison between the measured the mean energy variation $\Delta \beta\av{\Delta E }$ (blue circles) and the calculated mean heat flux $\Delta \beta\av{\Delta E }$ (orange diamonds). 
\textbf{(d-f)} Experimental verification of the quantum fluctuation relation $\av{e^{-\Delta \beta \Delta E }}=1$ [Eq.~(\ref{eq:FTb})]. 
The driving Hamiltonian varies in phase according to Eq.~(\ref{eq:H_fixed-amplitude}), with a period $\tau_\theta$. 
The system is also subject to short laser pulses with an interpulse spacing that coincides with the modulation period of the Hamiltonian. 
The constant value of $\Delta \beta \equiv \beta-\beta_R$ is defined by the probabilities $P_\uparrow{(0)}$ and $P_\uparrow^\infty$ to measure $E_\uparrow$ for the initial and asymptotic states, respectively. 
The probabilities associated with the initial states and the parameters defining the qubit dynamics are respectively 
\textbf{(a,d)} $P_\uparrow{(0)} =  0.509$, $\tau_\theta=1296\,\mathrm{ns}$, $\alpha=\pi/4$; \textbf{(b,d)} $P_\uparrow{(0)} =  0.303$, $\tau_\theta=616\,\mathrm{ns}$, $\alpha=\arctan(1/2)$; and \textbf{(c,f)} $P_\uparrow{(0)} =  0.126$, $\tau_\theta=308\,\mathrm{ns}$, $\alpha=\arctan(1/4)$. 
The values of $P_\uparrow^\infty$ are reported in Fig.~\ref{fig:condit_probs_phase}(b-d). 
\textbf{(d-f)} Experimental data is shown as blue circles while orange diamonds indicate the according theoretical prediction (i.e. the constant value $1$).
}
\label{fig:joint_probs_and_QHFR_phase} 
\end{figure}

In the further experiments, short laser pulses are applied at the discrete times $t_n=n\tau_\theta$, and final energy measurements are performed at a final time $t_{\mathrm{f}} = N_L \tau_\theta$. 
Accordingly, the spin free energy difference $\Delta F$ is null, and Eq.~(\ref{eq:FTb}) takes on the simpler form 
\begin{equation}
\av{e^{-(\beta-\beta_R) \Delta E }}= 1\, .
\label{eq:FTc}
\end{equation}

The experimental data are shown together with a Monte-Carlo simulation (see~\ref{app:num_model}). Figure \ref{fig:condit_probs_phase}(b-d) shows the impact of the laser pulses on the conditional probabilities $P_{\uparrow|\uparrow}$ and $P_{\uparrow|\downarrow}$ as a function of $N_L$. 
Note how these probabilities converge to the same asymptotic value $P_\uparrow^{\infty}$, meaning that, regardless of the initial condition, the system ends up in the $\ket{\uparrow}$ state with probability $P_\uparrow^{\infty}$. In other words, the system reaches a steady state which conforms to a Gibbs state of inverse temperature $\beta_{R} = -\frac{1}{\hbar\omega} \ln \left(\frac{P_{\uparrow}^{\infty}}{1-P_{\uparrow}^{\infty}}\right)$. 
The value of $P_{\uparrow}^{\infty}$, and therefore the value of $\beta_{R}$, depends on the ratio $\omega_0/\theta$. In the adiabatic regime, $\theta\ll\omega_0$, the Hamiltonian is approximately proportional to $\sigma_x$, and $P_{\uparrow}^{\infty}\simeq 1/2$. 
Hence, in the adiabatic regime the value of $\beta_{R}$ tends to zero. As the value of $\theta$ increases with respect to $\omega_0$, the dynamics no longer belongs to the adiabatic regime, and the value $P_{\uparrow}^{\infty}$ approaches to zero, hence increasing the value of $\beta_{R}$. 
Note that, although the spin is driven by a time dependent Hamiltonian that delivers power continuously, at discrete times in which the energy is measured the system evolves as under the influence of an effective thermal reservoir of inverse temperature $\beta_{R}$ and as if it were not in contact with any work source. 
Given these conditions, the mean energy variation is equal to the mean heat flux $\av{Q}$, as demonstrated in Fig.~\ref{fig:joint_probs_and_QHFR_phase}(a-c) by directly comparing $\av{\Delta E}$, extracted from the experimental data, with the mean heat flux calculated with the theoretical model. 
See~\ref{app:mean_W_and_Q} for details on how to calculate $\av{Q}$ in this case. 
Remarkably, even when $\av{\Delta E}$ is different from zero, we demonstrate that the energy variation statistics follows the quantum fluctuation relation $\av{e^{-(\beta-\beta_R) \Delta E }}= 1$~[Eq.~(\ref{eq:FTc})], as the general theory predicts.  
Fig.~\ref{fig:joint_probs_and_QHFR_phase}(d-f) shows the experimental values of the quantity $\av{e^{-(\beta-\beta_R) \Delta E }}$, as obtained under various experimental conditions (specifically with various values of $\tau_\theta$ and $\alpha$). The experimental data confirm the validity of the quantum fluctuation relation as expressed in Eq.~(\ref{eq:FTb}). 

\section{Conclusions}

The experimental verification of quantum fluctuation relations for open systems that simultaneously exchange heat---due to a thermal gradient---and work---due to a time-dependent driving---is a formidable challenge, owing to the general difficulty 
of measuring heat and work separately, while having typically experimental access to their difference $\Delta E$.

With this work we establish a first milestone along the way towards a fully fledged experimental demonstration of fluctuation relations for driven-dissipative systems. The verification becomes feasible ({\it i.e.}, it requires the statistics of $\Delta E$ only) in two special cases: {\it i)} the external environment is at infinite temperature, and {\it ii)} the system is subject to a periodic driving, so that, at multiples of the driving period the total work amounts to zero (while the power is generally non-null). In those cases the fluctuation relations boil down to the simplified expressions in Eq.~(\ref{eq:FTa}) and (\ref{eq:FTb}), respectively, which we have experimentally verified.

Our experimental platform of choice was a single NV center in diamond. Due to its long coherence time, and the possibility to finely control its state and its interaction with engineered environments, this system naturally lend itself to study non-equilibrium thermodynamics in the quantum regime \cite{HernandezGomez20}.

Our results represent the first experimental study of quantum fluctuation relation in driven open quantum systems.

\section*{Acknowledgments}
The authors would like to thank P. Cappellaro, F. Poggiali, and S. Gherardini for useful discussions, and F.S. Cataliotti and M. Inguscio for continuous support. 

\appendix

\section{Numerical model of the dynamics}
\label{app:num_model}

In order to model the qubit dynamics, we separate the evolution into a unitary part, describing the dynamics between consecutive short laser pulses, and the irreversible dynamics that describe the interaction with laser pulses.

The unitary operators of the map correspond to 
\begin{equation}
U(t_{N_\mathrm{L}-1},t_{N_\mathrm{L}})=\exp \left(-i \int_{t_{N_\mathrm{L}-1}}^{t_{N_\mathrm{L}}}\mathrm{d}t'\mathcal{H}(t') \right) ,
\end{equation}
with $t_N \equiv N \tau$. We recall that $\tau$ is the time between laser pulses. 
In the case of the Hamiltonian in Eq.~(\ref{eq:H_variable-amplitude}), considered in Sec.~\ref{sec:cyclicAmpVar}, each unitary operator describes a rotation around the $x$ axis, such that $U(t_{N_\mathrm{L}-1},t_{N_\mathrm{L}})=\exp \left(-i \sigma_x \int_{t_{N_\mathrm{L}-1}}^{t_{N_\mathrm{L}}}\mathrm{d}t'\omega(t') \right) $, with $\omega(t)$ defined by Eq.~(\ref{eq:omega_t}). 
In contrast, in the case studied in Sec.~\ref{sec:cyclicPhase}, the unitary evolution is described by the Floquet Hamiltonian $\mathcal{H}_{ \theta }=\frac{\hbar}{2} \left( \omega_0 \sigma_x + \theta \sigma_z \right)$~[Eq.~(\ref{eq:floquet_hamiltonian})] only for discrete times $t_n = n \tau_\theta$, such that, for interpulse times $\tau = \tau_\theta$, the unitary operator corresponds to $U(t_{N_\mathrm{L}-1},t_{N_\mathrm{L}})=U_\theta= \exp \left(-i \mathcal{H}_{ \theta } \tau_\theta / \hbar\right)$. 
Both of these unitary operators can be solved analytically and then included in a numerical model of the dynamics.

The dissipative dynamics associated with each laser pulse consists instead in an energy jump that is modeled as a POVM, with $(1-p_\mathrm{a})$ probability to leave the state unaffected, and a  probability $p_\mathrm{a}$ to absorb photons. Upon absorption, the spin qubit is subject to a projective measurement of $\sigma_z$ followed by an optical pumping of the spin modeled via Lindbladian master equation, with a Lindbladian jump operator $\sqrt{\Gamma}\ket{0}\!\!\bra{1}$ that transfers spin population to the state $\ket{0}$. 
The numerical simulation is then performed by using a Monte-Carlo simulation to determine the result of the POVM, hence determining if the laser pulse is absorbed or not. If the pulse is absorbed, Lindbladian operator is applied to the state of the system. In this way, we simulate single trajectories followed by the system, and we finally average over several realizations to obtain the mean trajectories shown in Fig.~\ref{fig:condit_probs_amplitude} and Fig.~\ref{fig:condit_probs_phase}. 

\section{Mean work and mean heat flux}
\label{app:mean_W_and_Q}

As described in the main text, the mean values of work and heat flux can be calculated as $\av{W} = \int_0^{t_\mathrm{f}} \mathrm{d}t' \mathrm{Tr} [\rho(t')\frac{\mathrm{d}\mathcal{H}(t')}{\mathrm{d}t'}]$, and $\av{Q} = \int_0^{t_\mathrm{f}} \mathrm{d}t' \mathrm{Tr} [\frac{\mathrm{d}\rho(t')}{\mathrm{d}t'} \mathcal{H}(t')]$~\cite{Alicki79}. In this Appendix we describe how to compute this quantities for the dynamics described in Sec.~\ref{sec:cyclicAmpVar} and Sec.~\ref{sec:cyclicPhase}. 

\subsection{Mean work and mean heat flux for Sec.~\ref{sec:cyclicAmpVar}}
\label{app:mean_W_and_Q_cyclicAmpVar}
The spin qubit is driven by the Hamiltonian~[Eq.~(\ref{eq:H_variable-amplitude})] $\mathcal{H}(t) = \frac{\hbar \omega(t)}{2} \sigma_x$, where $\omega(t)$ is a periodic function such that $\omega(t) = \omega(t + \tau_A) $. 
In addition the system interacts with short laser pulses applied at times $t_\mathrm{f}=N_\mathrm{L}\tau$, where $\tau$ is the time between consecutive pulses and $N_\mathrm{L}$ is the number of pulses. 
Assuming that the duration of every short laser pulse interaction is given by $\delta$, then the mean work and mean heat flux after the first laser pulse is given by 
\begin{eqnarray}
\av{W}_1 &=& \int_{0}^{\tau} \mathrm{d}t' \mathrm{Tr} [\rho(t')\frac{\mathrm{d}\mathcal{H}(t')}{\mathrm{d}t'}]  + \int_{\tau}^{\tau+\delta} \mathrm{d}t' \mathrm{Tr} [\rho(t')\frac{\mathrm{d}\mathcal{H}(\tau)}{\mathrm{d}t'}] \nonumber\\
&=& \int_{0}^{\tau} \mathrm{d}t' \mathrm{Tr} [\rho(t')\frac{\mathrm{d}\mathcal{H}(t')}{\mathrm{d}t'}] \label{eq:W1}
\end{eqnarray}
\begin{eqnarray}
\av{Q}_1 &=& \int_{0}^{\tau} \mathrm{d}t' \mathrm{Tr} [\frac{\mathrm{d}\rho(t')}{\mathrm{d}t'} \mathcal{H}(t')] + \int_{\tau}^{\tau+\delta} \mathrm{d}t' \mathrm{Tr} [\frac{\mathrm{d}\rho(t')}{\mathrm{d}t'} \mathcal{H}(\tau)] \nonumber\\
&=& \int_{\tau}^{\tau+\delta} \mathrm{d}t' \mathrm{Tr} [\frac{\mathrm{d}\rho(t')}{\mathrm{d}t'} \mathcal{H}(\tau)] \label{eq:Q1}
\end{eqnarray}
The second term in Eq.~(\ref{eq:W1}) vanishes because we assume the interaction with the laser pulses to be instantaneous, $\delta\ll\tau$ and $\delta\ll\tau_A$, hence the Hamiltonian is kept constant during the interaction with a single laser pulse.  
Instead, the first term in Eq.~(\ref{eq:Q1}) is equal to zero because the system evolves under unitary dynamics during the integration interval, meaning that $\frac{\mathrm{d}\rho}{\mathrm{d}t} = \frac{1}{i\hbar}[\mathcal{H},\rho]$. Then, the cyclic property of the trace implies that $\mathrm{Tr} [\frac{\mathrm{d}\rho}{\mathrm{d}t} \mathcal{H}] = \frac{1}{i\hbar}(\mathrm{Tr} [\mathcal{H}\rho\mathcal{H}] - \mathrm{Tr} [\rho\mathcal{H}^2])=0$. 
Therefore, for a given final time $t_\mathrm{f}$ such that $N_\mathrm{L}\tau < t_\mathrm{f}<(N_\mathrm{L}+1)\tau$ we can separate the mean work and heat flux as
\begin{equation} 
\av{W} = \sum_{n=1}^{N_\mathrm{L}} \av{W}_n + \int_{\tau N_\mathrm{L}}^{t_\mathrm{f}} \mathrm{d}t' \mathrm{Tr} [\rho(t')\frac{\mathrm{d}\mathcal{H}(t')}{\mathrm{d}t'}] \label{eq:W_all}
\end{equation}
\begin{equation}
\av{Q} = \sum_{n=1}^{N_\mathrm{L}} \av{Q}_n \label{eq:Q_all}
\end{equation}
where 
\begin{eqnarray}
\av{W}_n &\equiv& \int_{(n-1)\tau}^{n\tau} \mathrm{d}t' \mathrm{Tr} [\rho(t')\frac{\mathrm{d}\mathcal{H}(t')}{\mathrm{d}t'}] 
\nonumber\\
&=& \frac{\hbar}{2}\int_{(n-1)\tau}^{n\tau} \mathrm{d}t'  \frac{\mathrm{d}\omega(t')}{\mathrm{d}t'} \mathrm{Tr} [\rho(t')\sigma_x] \label{eq:W_n} 
\end{eqnarray}
\begin{eqnarray}
\av{Q}_n& \equiv& \int_{n\tau}^{n\tau+\delta} \mathrm{d}t' \mathrm{Tr} [\frac{\mathrm{d}\rho(t')}{\mathrm{d}t'} \mathcal{H}(\tau)] \nonumber\\
& = &\mathrm{Tr} [(\rho(n\tau+\delta)-\rho(n\tau)) \mathcal{H}(\tau)] . \label{eq:Q_n}
\end{eqnarray}
Note that the density operator at any given time $t$ can be written as 
\begin{equation}
\rho(t) =  P_+(t) \ket{E_{+}}\!\!\bra{E_{+}} + (1-P_+(t))\ket{E_{-}}\!\!\bra{E_{-}} + ( q(t) \ket{E_{-}}\!\!\bra{E_{+}} + \mathrm{c.c.} ) 
\end{equation} 
where $0\leq P_+(t)\leq 1$ is the population on the eigenstate $\ket{E_{+}}$, and $q(t)$ is a complex number describing the coherence of the state in the Hamiltonian basis. 
We recall that $\ket{E_{\pm}}$ are eigenstates of $\sigma_x$ with eigenvalues $\pm1$, which implies that $\mathrm{Tr} [\rho(t)\sigma_x] = 2P_+(t)-1$. 
Therefore, Eq.~(\ref{eq:Q_n}) can be rewritten as 
\begin{equation} \label{eq:Q_n_bis}
\av{Q}_n = \hbar\omega(n\tau) \Big( P_+(n\tau) - P_+((n-1)\tau) \Big). 
\end{equation} 
In addition, since the Hamiltonian is always parallel to $\sigma_x$, then the population $P_+(t)$ during the unitary evolution is kept constant. 
Therefore, Eq.~(\ref{eq:W_n}) can be recast as 
\begin{equation} \label{eq:W_n_bis}
\av{W}_n = \frac{\hbar}{2} \Big(2P_+((n-1)\tau)-1\Big) \Big(\omega(n\tau) - \omega((n-1)\tau) \Big)
\end{equation}
This means that knowing the population at times $t_\mathrm{f} = n \tau$ is enough to calculate $\av{W}_{n}$ and $\av{Q}_{n}$. The population, as we just mentioned, are unaltered by the unitary evolution, so we need to understand how they are affected by the short laser pulses. 
As described in~\ref{app:num_model}, the interaction with a single laser pulse has a probability $1-p_\mathrm{a}$ of leaving the system unaltered, and a probability $p_\mathrm{a}$ to project the state of the system into one of the eigenstates of $\sigma_z$, and then applying a Lindbladian evolution that \textit{pumps} population from $\ket{1}\!\!\bra{1}$ to $\ket{0}\!\!\bra{0}$. 
Hence, the state of the system after absorbing a laser pulse is diagonal in the $\sigma_z$ basis, meaning that its populations in the $\sigma_x$ basis must be equal to $1/2$.
Therefore, we can write a recursive equation for the population in the $\sigma_x$ basis 
\begin{equation} \label{eq:recursive_popul}
P_+(n\tau) = (1-p_\mathrm{a}) P_+((n-1)\tau) + p_\mathrm{a}/2 . 
\end{equation} 
Notice that $P_+(0)=1/(1+e)$, as mentioned in the main text.  
Therefore, Eq.~(\ref{eq:recursive_popul}) can be used recursively to obtain
\begin{equation} \label{eq:p_plus_nTau}
P_+(n\tau) = \frac{1}{2}\big( 1 - (1-p_\mathrm{a})^n (1-2P_+(0)) \big) . 
\end{equation} 

Finally, we can recast Eqs.~(\ref{eq:W_n_bis}) and (\ref{eq:Q_n_bis}) as
\begin{equation} 
\av{W}_n = \frac{\hbar}{2} \Big(\omega((n-1)\tau) - \omega(n\tau) \Big) (1-p_\mathrm{a})^{n-1} (1-2P_+(0)) \label{eq:W_n_tris}\end{equation}
\begin{equation}
\av{Q}_n = \frac{\hbar}{2} \omega(n\tau) p_\mathrm{a}(1-p_\mathrm{a})^{n-1} (1-2P_+(0)) . \label{eq:Q_n_tris}
\end{equation} 
Equations (\ref{eq:W_n_tris}) and (\ref{eq:Q_n_tris}) can be used to calculate the values of $\av{W}$ and $\av{Q}$~[Eqs.~(\ref{eq:W_all}-\ref{eq:Q_all})]. The mean values of work and heat flux are therefore completely defined by the parameters $N_\mathrm{L}$, $P_+(0)$, $t_\mathrm{f}$, $\tau/\tau_\mathrm{A}$, and $p_\mathrm{a}$, the latter being the only free parameter fitted to the experimental data in Fig.~\ref{fig:condit_probs_amplitude}.

\subsection{Mean heat flux for Sec.~\ref{sec:cyclicPhase}}
\label{app:mean_Q_cyclicPhase}
For the stroboscopic times described in Sec.~\ref{sec:cyclicPhase}, the unitary evolution of the system is described by $U_\theta= e^{-i \mathcal{H}_{ \theta } \tau_\theta / \hbar}$, where $\mathcal{H}_{ \theta }$ is the time independent Floquet Hamiltonian in Eq.~(\ref{eq:floquet_hamiltonian}), and $\tau_\theta$ is the time between two consecutive laser pulses, and the time at which the energy of the system is measured. 
Given that $\mathcal{H}_{ \theta }$ is time independent, then the mean work measured at stroboscopic times $t_\mathrm{f}=\tau N_\mathrm{L}$ is equal to zero. 
On the other hand, in analogy to~\ref{app:mean_W_and_Q_cyclicAmpVar}, the mean heat flux measured at those stroboscopic times is 
\begin{eqnarray}
\av{Q} &=& \sum_{n=1}^{N_\mathrm{L}} \av{Q}_n = \sum_{n=1}^{N_\mathrm{L}} \int_{n\tau}^{n\tau+\delta} \mathrm{d}t' \mathrm{Tr} [\frac{\mathrm{d}\rho(t')}{\mathrm{d}t'} \mathcal{H}_\theta] 
\nonumber \\
&=& \sum_{n=1}^{N_\mathrm{L}}  2 E_\theta \left( P_\uparrow(n\tau) - P_\uparrow((n-1)\tau) \right) 
\nonumber \\
  &= & 2 E_\theta \Big( P_\uparrow(N_\mathrm{L}\tau) - P_\uparrow(0) \Big),  \label{eq:Q_phase}
\end{eqnarray}
where $E_\theta = \hbar\sqrt{\omega_0^2 + \theta^2}/2$. 

Let us assume that the density operator $\rho(\tau)$ represents the state of the system after applying a single short laser pulse to the state $\rho(0)$. 
As explained in~\ref{app:num_model}, the qubit interaction with a short laser pulse is modeled as a POVM with probability $(1-p_\mathrm{a})$ to leave unaltered the state of the qubit and a probability $p_\mathrm{a}$ to apply a projective measurement of $\sigma_z$ followed by Lindbladian dynamics under the jump operator $\sqrt{\Gamma}\ket{0}\!\!\bra{1}$. 
Under this model, the probability to obtain $E_{\theta,\uparrow} = E_\theta$ when measuring the energy of $\rho(\tau)$ is
\begin{eqnarray}
P_\uparrow(\tau) &=& \mathrm{Tr} [ \ket{E_{\theta,\uparrow}}\!\!\bra{E_{\theta,\uparrow}} \rho(\tau)]
\nonumber\\
&=& (1 - p_\mathrm{a} k) P_\uparrow(0) +  \frac{1}{2} p_\mathrm{a} (k - p_\mathrm{d} \cos \alpha) \label{eq:recursive_popul_phase}
\end{eqnarray}
where $p_\mathrm{d} \equiv 1 - e^{-\Gamma \delta}$ is the optical pumping probability associated with the Lindbladian dynamics, $\alpha = -\arctan\left({\omega_0}/{\theta}\right)$ is the angle that defines the ratio between the terms in the Floquet Hamiltonian~[Eq.~(\ref{eq:floquet_hamiltonian})], $P_\uparrow(0) = \mathrm{Tr} [ \ket{E_{\theta,\uparrow}}\!\!\bra{E_{\theta,\uparrow}} \rho(0)]$ is the probability associated with the initial state, and where we have defined, to simplify notation, a factor
\begin{equation}
k \equiv 1 + (1 - p_\mathrm{d}) \cos\alpha^2 .
\end{equation} 
The relation in Eq.~(\ref{eq:recursive_popul_phase}) can be recursively applied in order to obtain 
\begin{equation}
P_\uparrow(N_\mathrm{L}\tau) = (1 - p_\mathrm{a} k)^{N_\mathrm{L}} P_\uparrow(0) + \frac{1}{2}(1-(1 - p_\mathrm{a} k)^{N_\mathrm{L}}) (1 - \frac{p_\mathrm{d}}{k} \cos \alpha)
\end{equation}
Therefore, the mean heat flux in Eq.~(\ref{eq:Q_phase}) can be recast as 
\begin{equation}
\langle Q \rangle = E_\theta \left(1-(1 - p_\mathrm{a} k)^{N_\mathrm{L}}\right) \left( 1 - \frac{p_\mathrm{d}}{k} \cos \alpha  - 2 P_\uparrow(0) \right).
\label{eq:Qphase2}
\end{equation}
While  $p_\mathrm{a}$ and $p_\mathrm{d}$ depend on the laser pulse duration and intensity, all the other parameters are precisely set in experiment, with negligible uncertainty. Note that $p_\mathrm{d}$ can be extracted from the asymptotic state probability $P_\uparrow^\infty = \lim_{N_\mathrm{L}\rightarrow\infty}P_\uparrow(N_\mathrm{L}\tau) = \frac{1}{2}(1 - \frac{p_\mathrm{d}}{k} \cos \alpha)$. Thus, the only free parameter in Eq.~(\ref{eq:Qphase2}) is the laser absorption probability $p_\mathrm{a}$, which we fit to the experimental data (see Fig.~\ref{fig:condit_probs_phase}). 

\section*{References}
\bibliographystyle{iopart-num-mod}
\bibliography{../OQS-Biblio}

\end{document}